\documentclass[reprint,showpacs,amsmath,amssymb,aps,prb]{revtex4-1}

\usepackage{graphicx}	
\usepackage{dcolumn}	
\usepackage{bm}			
\usepackage{here}

\begin{document}

\preprint{APS/123-QED}

\title{Superconductivity and the electronic phase diagram of LaPt$_{2-x}$Ge$_{2+x}$}

\author{S. Maeda$^1$}
\author{K. Matano$^1$}
\author{R. Yatagai$^1$}
\author{T. Oguchi$^2$}
\author{Guo-qing Zheng$^{1,3}$}

\affiliation{$^1$Department of Physics, Okayama University, Okayama 700-8530, Japan}
\affiliation{$^2$Institute of Scientific and Industrial Research, Osaka University, Ibaraki, Osaka 567-0047, Japan}
\affiliation{$^3$Institute of Physics and Beijing National Laboratory for Condensed Matter Physics, Chinese Academy of Sciences, Beijing 100190, China}



\begin{abstract}
In many cases, unconventional superconductivity are realized by suppressing another order parameter, such as charge density wave (CDW) or spin density wave (SDW).
This suggests that the fluctuations of these order parameters play an important role in producing superconductivity.
LaPt$_2$Ge$_2$ undergoes a tetragonal-to-monoclinic structural phase transition (SPT) at $T_{\rm s}$ = 394 K, accompanying a double period modulation in the $a$-axis direction, and superconducts at $T_{\rm c}$ = 0.41 K.
We performed band calculations  and found 2D (two dimensional)-like Fermi surfaces with partial nesting.
A reduction in the density of states in the monoclinic phase  was found in the calculation and confirmed  by $^{195}$Pt-NMR. We suggest a CDW as a possible cause for the SPT.
By changing the stoichiometry between Pt and Ge, we succeeded in suppressing $T_{\rm s}$ and increasing $T_{\rm c}$ in LaPt$_{2-x}$Ge$_{2+x}$.
Comparison of $^{139}$La- and $^{195}$Pt-NMR data reveals moderate fluctuations associated with SPT.
From $^{139}$La-NQR measurements at zero field, we found that an isotropic superconducting gap is realized in LaPt$_{2-x}$Ge$_{2+x}$ ($x$ = 0.20).
We discuss the relationship between superconductivity and the SPT order/fluctuations.
\end{abstract}


\maketitle

\section{Introduction}
\begin{figure}[h]
\begin{center}
\includegraphics[width=8cm]{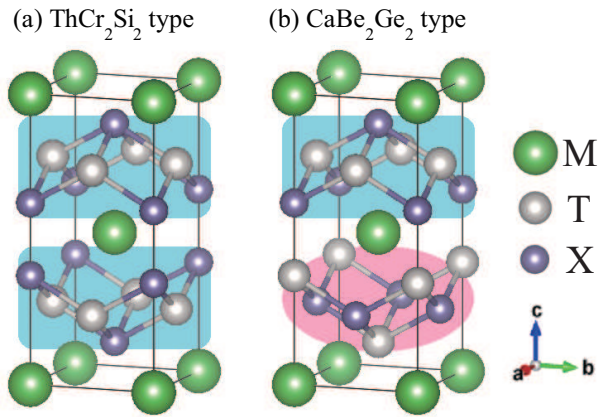}
\caption{(Color online) Crystal structure of the ThCr$_2$Si$_2$ type (a) and CaBe$_2$Ge$_2$ type (b). The blue rectangle and the pink oval represent the [X-T-X] and the [T-X-T] layers, respectively.}
\label{MT2X2}
\end{center}
\end{figure}

Most ternary compounds in the formula of MT$_2$X$_2$ (M = rare earth or alkaline earth metals, T = transition metals, X = Si or Ge), including some Fe-based superconductors, crystallize in the body centered tetragonal ThCr$_2$Si$_2$ type structure\cite{Parthe} (Fig. \ref{MT2X2}(a)).
On the other hand, some MT$_2$X$_2$ with T = Ir or Pt crystallize in the primitive tetragonal CaBe$_2$Ge$_2$ type structure\cite{Shelton,Dommann} (Fig. \ref{MT2X2}(b)).
The CaBe$_2$Ge$_2$ type structure is closely related to the ThCr$_2$Si$_2$ type structure.
The ThCr$_2$Si$_2$ type structure has two [X-T-X] layers along the $c$ axis while in the CaBe$_2$Ge$_2$ type structure, one of them is replaced by [T-X-T] layer.
Some of non-magnetic CaBe$_2$Ge$_2$ type compounds (SrPt$_2$As$_2$\cite{Imre2,SrPt2As2_1,SPALPSFS,Kawasaki} and LaPt$_2$Si$_2$\cite{LaPt(2)Si2,LPSNMR,SPALPSFS}) show a coexistence of superconductivity and charge density wave (CDW).
In these compounds, the CDW occurs in the [X-T-X] layers.

Fluctuations associated with CDW or spin density wave (SDW) have attracted much attention in recent years, since they may be responsible for superconductivity that is realized near a CDW or SDW phase.
In the high-$T_{\rm c}$ cuprates, spin fluctuations arising from the nearby antiferromagnetism have been studied extensively.
Recently, CDW has also attracted attention in relation to the unusual normal state in the cuprates\cite{cuprate}.
In the Fe pnictides, both spin fluctuations and orbital or structural fluctuations are believed to be important.
For example, underdoped BaFe$_{2-x}$M$_x$As$_2$ (M = Co, Ni) are metallic and show a tetragonal-to-orthorhombic structural phase transition (SPT) followed by an SDW order\cite{CoDoped,Zhou}.
Superconductivity appears after suppressing these orders, and fluctuations associated with SPT and SDW have been observed\cite{Zhou}.
Compounds such as SrPt$_2$As$_2$ and LaPt$_2$Si$_2$ showing a coexistence of superconductivity and CDW can be recognized as non-magnetic versions of Fe-based superconductors, because both CDW and SDW originate from a Fermi surface nesting and the crystal structure is very similar.

\begin{figure}[h]
\begin{center}
\includegraphics[width=8cm]{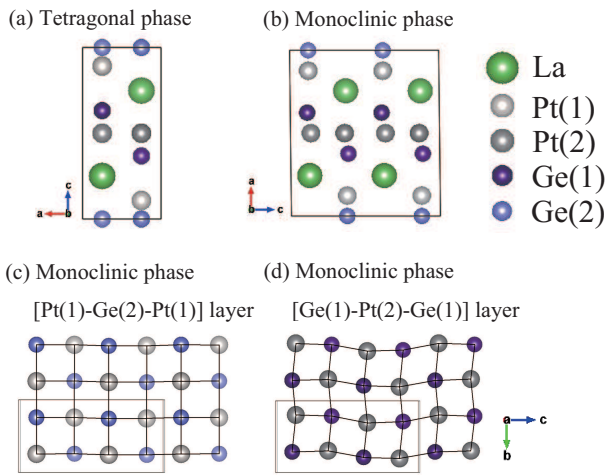}
\caption{(Color online) Crystal structure of the tetragonal (a) and monoclinic (b) phases, [Pt(1)-Ge(2)-Pt(1)] (c) and [Ge(1)-Pt(2)-Ge(1)] (d) layers of the monoclinic phase. The solid rectangle represents the unit cell.}
\label{CrystalStructure}
\end{center}
\end{figure}

LaPt$_2$Ge$_2$ is a superconductor with $T_{\rm c}$ = 0.41 K\cite{Hull,Maeda} and shows a SPT at $T_{\rm s}$ = 385.8 K\cite{Imre}.
As shown in Figs. \ref{CrystalStructure}(a) and \ref{CrystalStructure}(b), the crystal structure of the high temperature phase is a tetragonal CaBe$_2$Ge$_2$ type (space group: $P4/nmm$), while that of the low temperature phase is a monoclinically distorted CaBe$_2$Ge$_2$ type ($P2_1/c$)\cite{Imre,Maeda}.
The monoclinic phase has a doubled unit cell in the $a$-axis direction.
As shown in Figs. \ref{CrystalStructure}(c) and \ref{CrystalStructure}(d), the monoclinic distortion is mainly in the [Ge(1)-Pt(2)-Ge(1)] layer, while there is almost no distortion in the [Pt(1)-Ge(2)-Pt(1)] layer.
The origin of the SPT, and its relationship to superconductivity is unknown.

In this work, we address this issue by band calculations, material synthesis, and nuclear magnetic resonance (NMR) measurements.
Our results suggest that CDW is a possible origin for the SPT. 
By suppressing the SPT, the density of states at the Fermi level is increased and $T_{\rm c}$ is enhanced.
We have also performed nuclear quadrupole resonance (NQR) measurements to study the superconducting gap.

\section{Methods}
The full relativistic band calculations including spin-orbit coupling (SOC) were performed with the all-electron full-potential linear augmented plane wave (FLAPW) method implemented in HiLAPW\cite{HiLAPW}.
Details of numerical procedures were described in the previous work\cite{HiLAPW}.
Polycrystalline samples of LaPt$_{2-x}$Ge$_{2+x}$ were synthesized by melting the elements of La (99.9 \%), Pt (99.999 \%), and Ge (99.999 \%) in an arc furnace under high purity (99.9999 \%) Ar atmosphere.
The resultant ingot was turned over and re-melted several times to ensure good homogeneity.
The weight loss during the arc melting was less than 1\%.
Subsequently, the samples were wrapped in Ta foil, sealed in a quartz tube filled with He gas,
annealed at 1000 $^\circ$C for 3 days and then slowly cooled to room temperature over a period of 3 days.
The samples were characterized by powder X-ray diffraction using Rigaku RINT-TTR III at room temperature.
The XRD patterns were analyzed by the RIETAN-FP program\cite{RIETAN}.
The crystal structure is drawn by using the VESTA program\cite{VESTA}.
The resistivity was measured by using a dc four-terminal method in the temperature range of 1.4 $\sim$ 480 K.
For ac susceptibility and NMR/NQR measurements, a part of the ingot was powdered.
The $T_{\rm c}$ was determined as the onset of the diamagnetism obtained by measuring the inductance of a coil filled with a sample which is a typical setup for NMR/NQR measurements.
The $T_{\rm{s}}$ was determined by two ways: by the minimum of the electrical resistivity ($x \leq  0.1$) or by the maximum of the $^{139}$La-NMR $1/T_1T$ ($x$ = 0.20).
The measurements below 1.4 K were carried out with a $^3$He-$^4$He dilution refrigerator.
NMR/NQR were carried out by using a phase-coherent spectrometer.
The NMR spectrum was obtained by integrating the spin echo intensity by changing the resonance frequency ($f$) at the fixed magnetic field of $H_0$ = 12.951 T.
The Knight shift ($K$) was determined by $K = (f_{\rm{peak}} - \gamma _{\rm{N}}H_0)/\gamma _{\rm{N}}H_0$, where $f_{\rm{peak}}$ is the peak frequency and $\gamma _{\rm{N}}$ = 9.094 MHz/T for $^{195}$Pt is the nuclear magnetic ratio.
The spin-lattice relaxation rate ($1/T_1$) was measured by using a single saturating pulse, and determined by fitting the recovery curve of the nuclear magnetization to the theoretical function\cite{Gordon,NQRRC}: $(M_0-M(t))/M_0 = \exp(-t/T_1)$ for $^{195}$Pt-NMR, $(M_0-M(t))/M_0 = (1/84)\exp(-t/T_1)+(3/44)\exp(-6t/T_1)+(75/364)\exp(-15t/T_1)+(1225/1716)\exp(-28t/T_1)$ for $^{139}$La-NMR (center peak) and $(M_0-M(t))/M_0 = 0.076994 \exp(-3t/T_1)+0.016497 \exp(-8.561749t/T_1)+0.906509 \exp(-17.206772t/T_1)$ for $^{139}$La-NQR ($\eta = 0.46$, $m = \pm 3/2 \leftrightarrow \pm 5/2$ transition), where $M_0$ and $M(t)$ are the nuclear magnetization in the thermal equilibrium and at a time $t$ after the saturating pulse.

\section{Results}
\subsection{Electronic structure calculations}
\begin{figure}[h]
\begin{center}
\includegraphics[width=7cm]{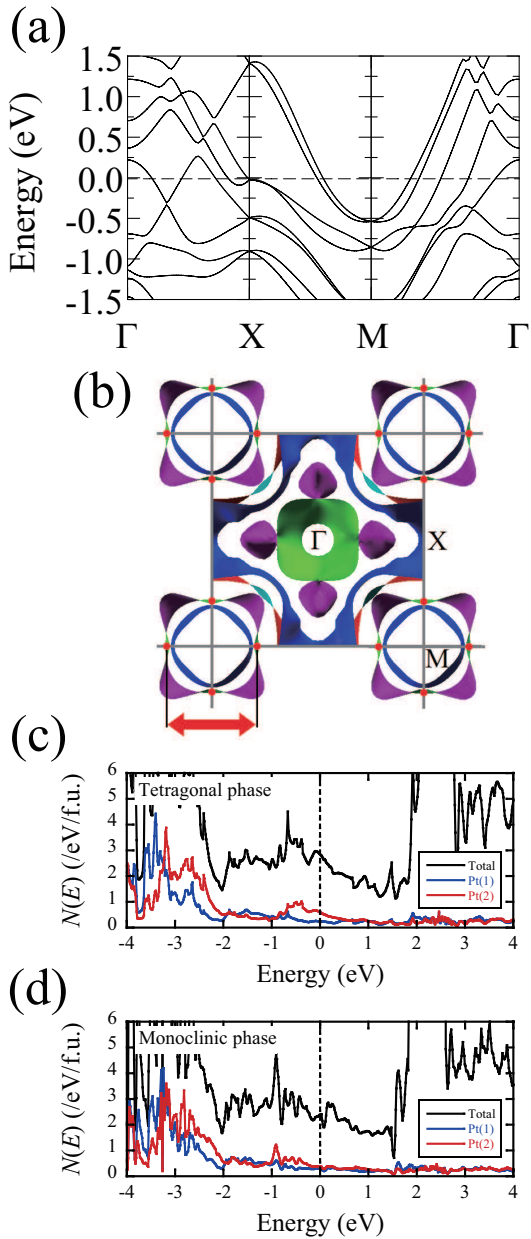}
\caption{(Color online) Band structure (a) and Fermi surface (b) of the tetragonal phase. The red arrows represent a nesting vector. Total and partial $N(E)$ of the tetragonal (c) and monoclinic (d) phases. The Fermi level is taken at the origin in (a), (c), and (d).}
\label{band}
\end{center}
\end{figure}

In the band calculation, we used the crystal structure determined for single crystal LaPt$_2$Ge$_2$ by Imre $et\ al.$\cite{Imre}.
Figures \ref{band}(a) and \ref{band}(b) show the band structure and the Fermi surface of the tetragonal phase, respectively.
These results are very similar to those of LaPt$_2$Si$_2$\cite{SPALPSFS}.
The Fermi surface consists of five sheets, and there are two 2D-like sheets around $M$ point.
The outer Fermi surface shows a partial nesting, similar to the case of LaPt$_2$Si$_2$\cite{SPALPSFS}.
Figures \ref{band}(c) and \ref{band}(d) display the total and partial density of states ($N(E)$), respectively.
Comparison between the tetragonal and monoclinic phases shows that the total $N(E)$ at the Fermi level ($E_{\rm F}$) of the monoclinic phase is 18 \% smaller than that of the tetragonal phase.
The characteristics of the calculated  Fermi surface usually favor the formation of a CDW state and the reduction of the $N(E_{\rm F})$ can be considered as a consequence of a CDW formation.
Therefore, the SPT of LaPt$_2$Ge$_2$ is possibly due to a commensurate CDW, although future measurements such as superlattice reflection are required to directly confirm this.
Looking in detail, the partial $N(E_{\rm F})$ at Pt(1) for the tetragonal and monoclinic phases are very similar.
The partial $N(E_{\rm F})$ at Pt(2) is larger than that at Pt(1) in the tetragonal phase, while it is reduced to a value almost the same as that at Pt(1) in the monoclinic phase.
These results are consistent with the change in the crystal structure.

\subsection{Basic physical properties of LaPt$_{2-x}$Ge$_{2+x}$}
\begin{figure}[h]
\begin{center}
\includegraphics[width=8cm]{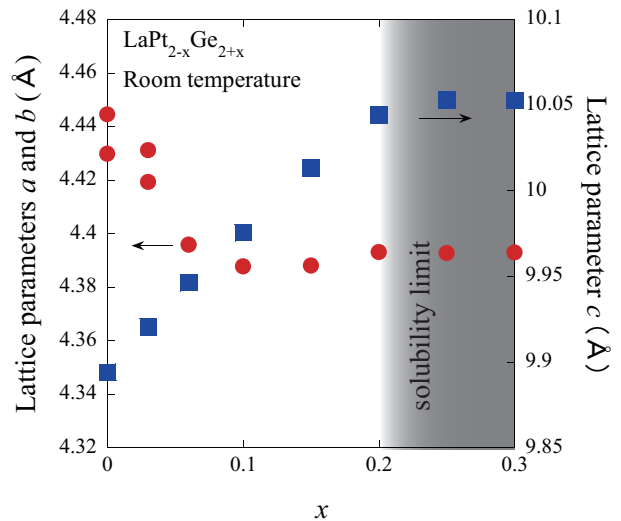}
\caption{(Color online) $x$ dependence of the lattice parameters $a$, $b$ and $c$ for LaPt$_{2-x}$Ge$_{2+x}$.The monoclinic structures are interpreted in the notation of tetragonal structure.}
\label{Lattice_parameters}
\end{center}
\end{figure}

No impurity peaks were observed in the XRD pattern in the range of 0 $\le$ $x$ $\le$ 0.30.
The extra Ge in non-stoichiometric LaPt$_{2-x}$Ge$_{2+x}$ ($x > 0$) is assumed to occupy deficient Pt sites because Pt and Ge sites are equivalent in the CaBe$_2$Ge$_2$ type structure.
Figure \ref{Lattice_parameters} shows the $x$ dependence of the lattice parameters for LaPt$_{2-x}$Ge$_{2+x}$ at room temperature.
With increasing $x$, the $c$-axis length increases linearly up to $x$ = 0.20.
Beyond $x$ = 0.20, the $c$-axis length is saturated, which suggests that the solubility limit is $x$ = 0.20.
On the other hand, the length $a$ and $b$ decrease.
For $x \geq 0.06$, $a$ and $b$ become constant since the compounds are in the tetragonal structure.

\begin{figure}[h]
\begin{center}
\includegraphics[width=8cm]{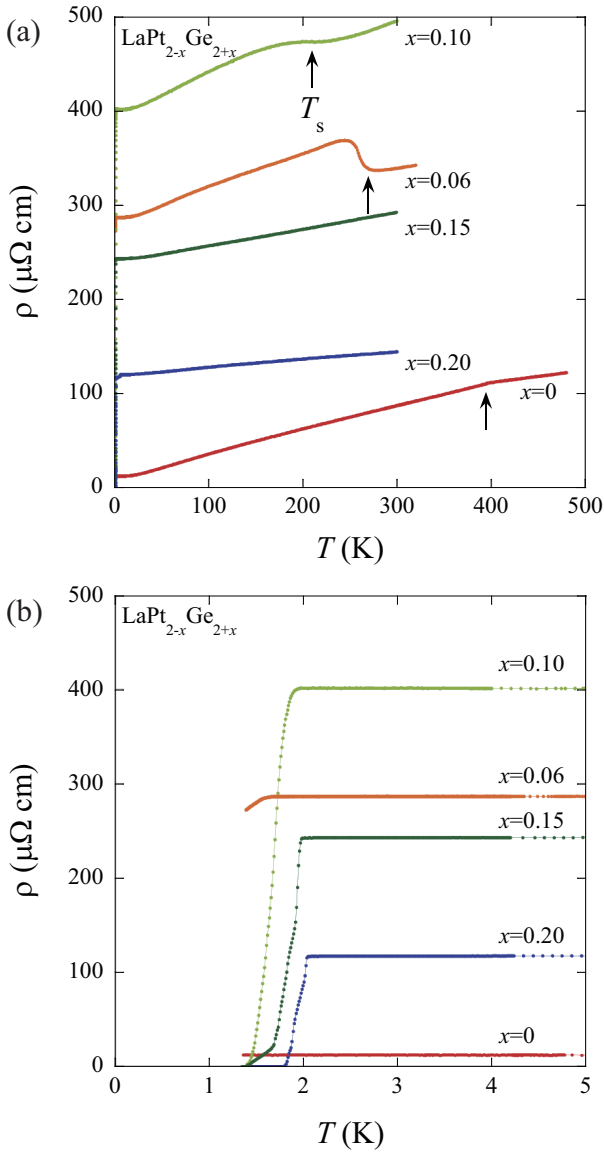}
\caption{(Color online) (a) Temperature dependence of the electrical resistivity for LaPt$_{2-x}$Ge$_{2+x}$. The solid arrows indicate $T_{\rm{s}}$. (b) Magnified view of the low temperature range.}
\label{rho}
\end{center}
\end{figure}

Figure \ref{rho}(a) shows the temperature dependence of the electrical resistivity for LaPt$_{2-x}$Ge$_{2+x}$.
The electrical resistivity for $x$ = 0 showed a kink at $T_{\rm{s}}$ = 394 K due to the SPT.
This is in good agreement with the value of $T_{\rm{s}}$ = 385.8 K reported by Imre $et\ al.$\cite{Imre}.
With increasing $x$, $T_{\rm{s}}$ decreased.
The strong anomaly for $x$ = 0.06 is because the sample cracked due to the SPT.
For $x$ = 0.15, the anomaly due to the SPT disappeared.
Figure \ref{rho}(b) shows the magnified view of the low temperature range.
The $T_{\rm c}$ increased with increasing $x$, with the highest $T_{\rm c}$ for $x$ = 0.20.

\begin{figure}[h]
\begin{center}
\includegraphics[width=8cm]{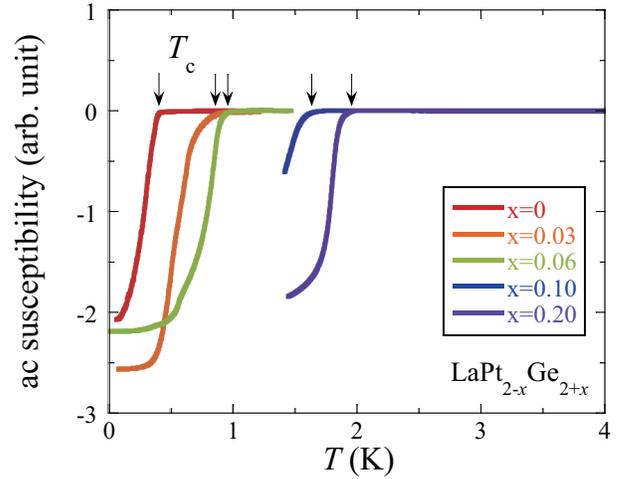}
\caption{(Color online) Temperature dependence of the ac susceptibility for LaPt$_{2-x}$Ge$_{2+x}$. The solid arrows indicate $T_{\rm c}$.}
\label{chi}
\end{center}
\end{figure}

Figure \ref{chi} shows the temperature dependence of the ac susceptibility measured by the NMR/NQR coil for LaPt$_{2-x}$Ge$_{2+x}$.
All samples showed a decrease in the ac susceptibility below $T_{\rm c}$, which is consistent with the electrical resistivity.

\subsection{Fluctuations and electronic state change due to structural phase transition}
We performed both $^{195}$Pt- and $^{139}$La-NMR measurements for the $x$ = 0.06 and $x$ = 0.20 samples at a fixed magnetic field of 12.951 T.
Because $^{139}$La has nuclear spin 7/2, fluctuations due to both the hyperfine field and the electric field gradient can be probed by the spin-lattice relaxation.
By contrast, $^{195}$Pt-NMR can only see the former, since $^{195}$Pt has nuclear spin 1/2.

\begin{figure}[h]
\begin{center}
\includegraphics[width=8cm]{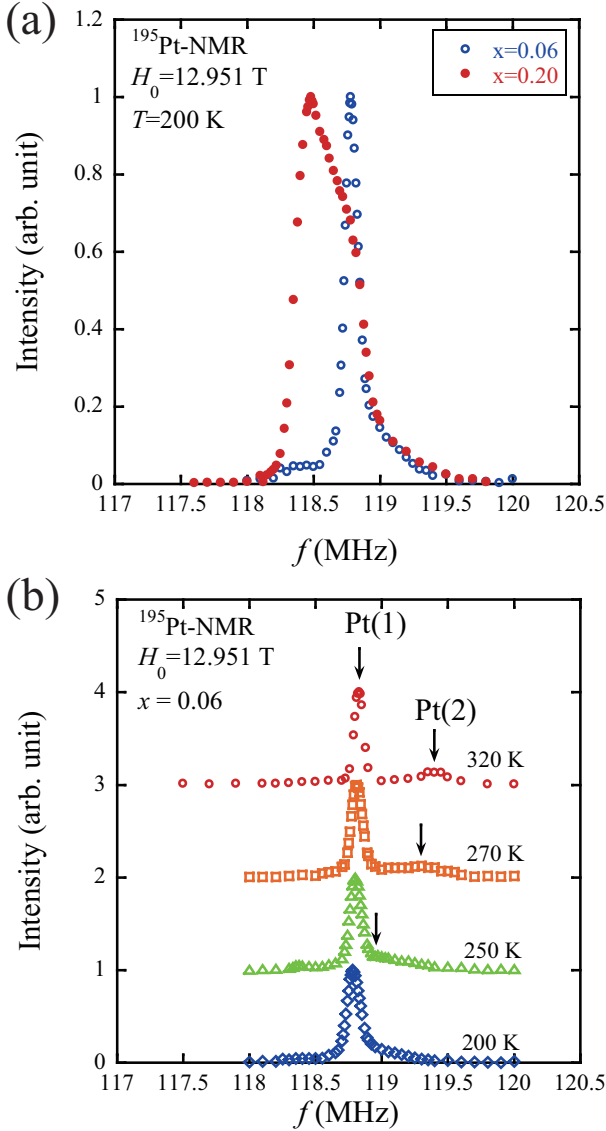}
\caption{(Color online) (a) $^{195}$Pt-NMR spectra for LaPt$_{2-x}$Ge$_{2+x}$ ($x = 0.06$ and 0.20) at 200 K. (b) $^{195}$Pt-NMR spectra for LaPt$_{2-x}$Ge$_{2+x}$ ($x = 0.06$) at various temperatures.}
\label{Pt_Spectra}
\end{center}
\end{figure}

Figure \ref{Pt_Spectra}(a) shows the $^{195}$Pt-NMR spectra at $T$ = 200 K.
The spectrum of $x$ = 0.06 is sharper than that of $x$ = 0.20, probably because the grains in $x$ = 0.06 are well orientated to the magnetic field towards the high susceptibility direction.
The grains in $x$ = 0.20 are orientated only partially, resulting in spectrum close to a powder pattern.
The degree of orientation depends on the anisotropy of the susceptibility, the size of the domains and the size of the grains.
We speculate that the domains of $x$ = 0.06 have grown much larger than that of $x$ = 0.20 because $x$ = 0.06 is closer to the stoichiometry. 

The spectrum of $x$ = 0.06 shows two peaks at high temperatures (Fig. \ref{Pt_Spectra}(b)).
This is because there are two Pt sites in LaPt$_2$Ge$_2$.
Since Pt(2) is mainly affected by the SPT, we identified the high frequency peak having large $T$-dependence as Pt(2), and the low frequency peak as Pt(1).
For $x$ = 0.20, two Pt sites cannot be distinguished because of the powder pattern.

\begin{figure}[h]
\begin{center}
\includegraphics[width=8cm]{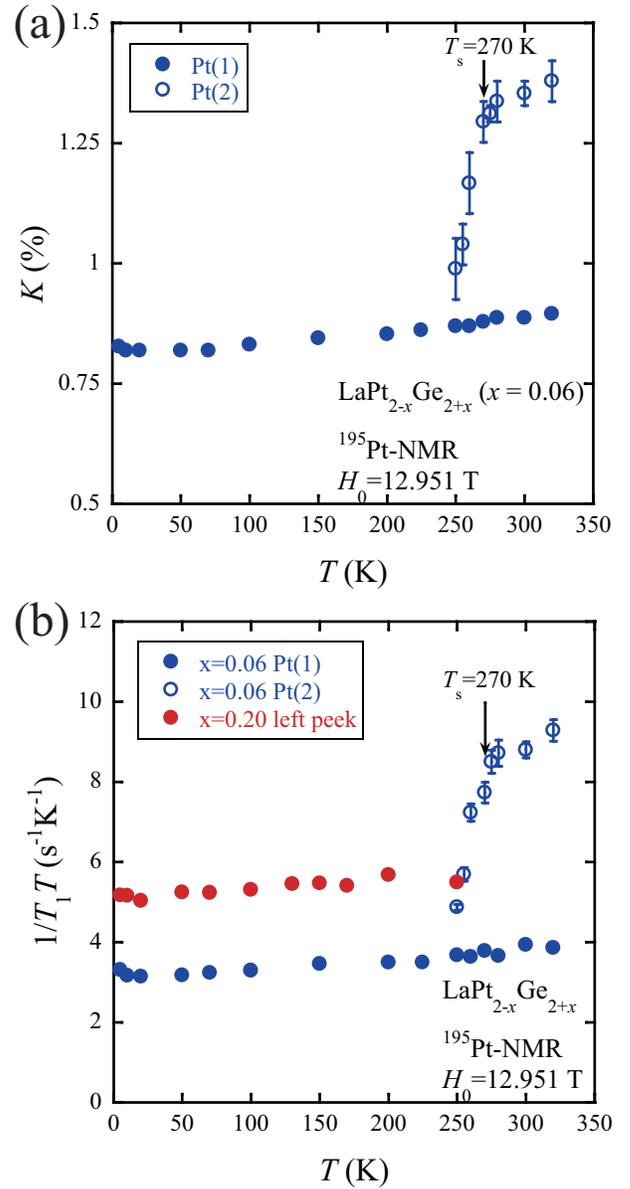}
\caption{(Color online) (a) Temperature dependence of the $^{195}$Pt-NMR Knight shift for LaPt$_{2-x}$Ge$_{2+x}$ ($x$ = 0.06). (b) Temperature dependence of the $^{195}$Pt-NMR $1/T_1T$ for LaPt$_{2-x}$Ge$_{2+x}$ ($x$ = 0.06 and 0.20).}
\label{PtNMR}
\end{center}
\end{figure}

The temperature dependence of the Knight shift ($K$) obtained from these spectra is shown in Fig. \ref{PtNMR}(a).
The $K$ reflects $N(E_{\rm F})$, through $K = K_{\rm o} + K_{\rm s}$ and $K_{\rm s} = A\mu_{\rm B}N(E_{\rm F})$.
Here $K_{\rm o}$ and $K_{\rm s}$ are $T$-independent orbital part and $T$-dependent spin part, respectively, and $A$ and $\mu_{\rm B}$ are the hyperfine coupling constant and the Bohr magneton, respectively.
For Pt(2), a large decrease in the $K$ due to the change in the $N(E_{\rm F})$ was observed around $T_{\rm{s}}$, while for Pt(1), no change due to the SPT was observed.

The quantity $1/T_1T$ also reflects the $N(E_{\rm F})$ through the relation $1/T_1T = A^2\pi {k_{\rm B}}\gamma_{\rm n}^2\hbar N^2(E_{\rm F}) + (1/T_1T)_{\rm{F}}$.
Here, the $T$-dependent $(1/T_1T)_{\rm{F}}$ is due to magnetic or electric fluctuations whose frequency is equal to the NMR frequency $f$.
For $x$ = 0.06, the temperature dependence of $1/T_1T$ (Fig. \ref{PtNMR}(b)) and $K$ is consistent.
For Pt(2), the $1/T_1T$ decreased around $T_{\rm{s}}$ because of the decrease in the $N(E_{\rm F})$ and became almost the same as that of Pt(1).
While for the Pt(1) site, no change due to the SPT was observed.
These results are consistent with the band calculations.
For $x$ = 0.20 where it is difficult to distinguish Pt(1) and Pt(2) in the spectrum (Fig. \ref{Pt_Spectra}), the $T_1$ was measured at the left peak where the Pt(1) signal is dominant.
No clear change was observed in the temperature dependence of the $1/T_1T$.

\begin{figure}[h]
\begin{center}
\includegraphics[width=8cm]{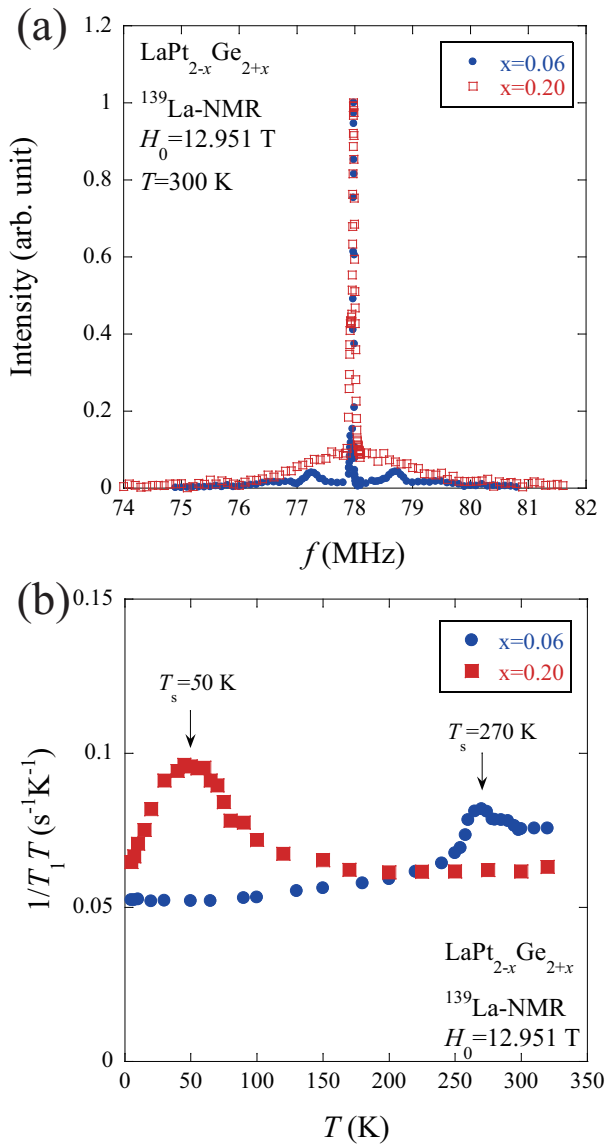}
\caption{(Color online) (a)$^{139}$La-NMR spectra for LaPt$_{2-x}$Ge$_{2+x}$ ($x$ = 0.06 and 0.20) at 300 K. (b)Temperature dependence of the $^{139}$La-NMR $1/T_1T$ for LaPt$_{2-x}$Ge$_{2+x}$ ($x$ = 0.06 and 0.20).}
\label{LaNMR}
\end{center}
\end{figure}

Figure \ref{LaNMR}(a) shows $^{139}$La-NMR spectra at $T$ = 300 K.
For $x$ = 0.06, distinct satellite peaks were observed due to orientation.
While for $x$ = 0.20, a broad powder pattern was observed, which is consistent with $^{195}$Pt-NMR spectra.
The $1/T_1T$ measured at the center peak is shown in Fig. \ref{LaNMR}(b).
For $x$ = 0.06, the $1/T_1T$ increased upon cooling to $T_{\rm{s}}$, and then decreased rapidly because of the SPT.
For $x$ = 0.20, similar behavior was observed although the resistivity showed no anomaly.
We determined the $T_{\rm{s}}$ at 50 K for $x$ = 0.20 by the maximum of the $^{139}$La-NMR $1/T_1T$.
These results are in sharp contrast with those for $^{195}$Pt-NMR.
Such upturn in $^{139}$La-NMR $1/T_1T$ can be understood as due to fluctuations associated with SPT that couple to the electric quadrupole moment of $^{139}$La nuclei.

\subsection{Phase diagram for LaPt$_{2-x}$Ge$_{2+x}$ and the superconducting gap probed by $^{139}$La-NQR}
\begin{figure}[h]
\begin{center}
\includegraphics[width=8cm]{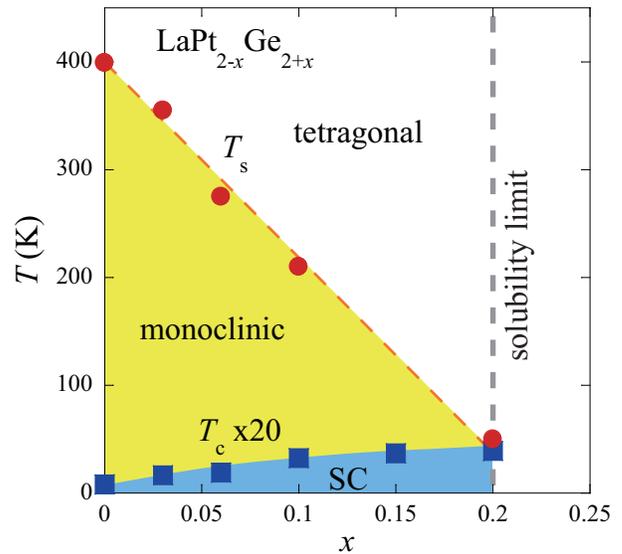}
\caption{(Color online) Phase diagram for LaPt$_{2-x}$Ge$_{2+x}$ obtained by the present work.
The red circles represent the $T_{\rm{s}}$ determined by a minimum of the electrical resistivity ($x < 0.20$) or a maximum of the $^{139}$La-NMR $1/T_1T$ ($x = 0.20$).
The blue squares represent the $T_{\rm c}$ determined as the onset temperature of the ac susceptibility.
The $T_{\rm c}$ is multiplied by 20 for clarity.}
\label{diagram}
\end{center}
\end{figure}

Figure \ref{diagram} shows the phase diagram for LaPt$_{2-x}$Ge$_{2+x}$ obtained in the present work.
The $T_{\rm s}$ decreased linearly and the $T_{\rm c}$ increased with increasing $x$.
The maximum $T_{\rm c}$ was 1.95 K in $x$ = 0.20.
The SPT critical point ($T_{\rm{s}} = 0$) is estimated to be $x_{\rm c} = 0.22$ by extrapolating the data for $x \leq 0.2$.
It is notable that the lattice parameter $c$ simultaneously increases with increasing $x$ (Fig. \ref{Lattice_parameters}), which suggests that weakening of the interaction between [Pt(1)-Ge(2)-Pt(1)] and [Ge(1)-Pt(2)-Ge(1)] layers may lead to the suppression of  the $T_{s}$.

\begin{figure}[h]
\begin{center}
\includegraphics[width=6cm]{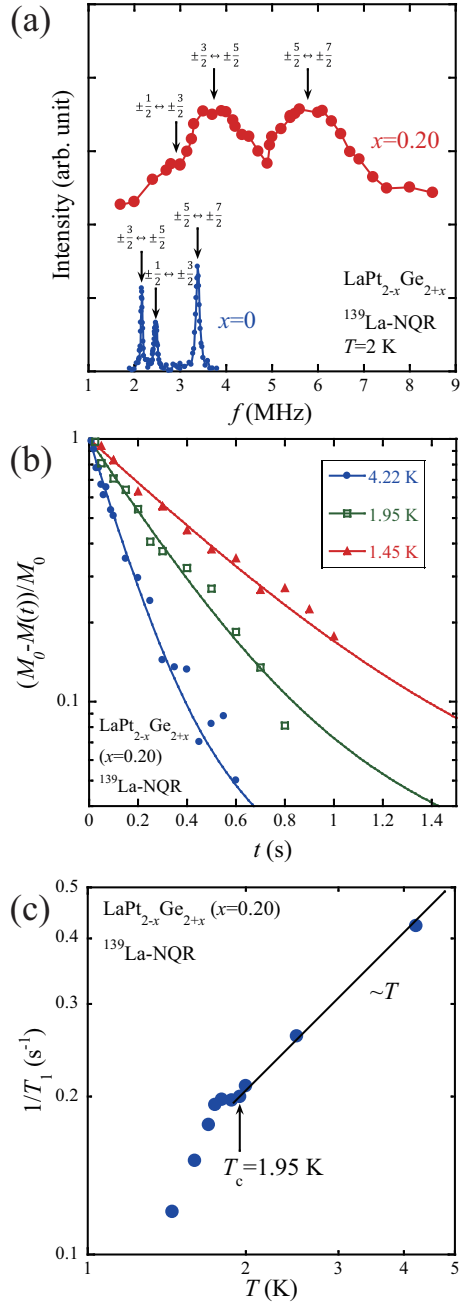}
\caption{(Color online) (a)$^{139}$La-NQR spectra for LaPt$_{2-x}$Ge$_{2+x}$ ($x$ = 0 and 0.20). (b)Recovery curves for LaPt$_{2-x}$Ge$_{2+x}$ ($x$ = 0.20) measured at the $m = \pm 3/2 \leftrightarrow \pm 5/2$ transition. (c)Temperature dependence of the $1/T_{1}$ for LaPt$_{2-x}$Ge$_{2+x}$ ($x = 0.20$).}
\label{NQR}
\end{center}
\end{figure}

Since LaPt$_{2-x}$Ge$_{2+x}$ has a low $H_{\rm{c2}}$ ($<$ 0.5 T), measurements with a magnetic field applied, such as the Knight shift in the superconducting state, are difficult.
In this work, we performed $^{139}$La-NQR at zero field to study the energy gap in the superconducting state.
Figure \ref{NQR}(a) shows the $^{139}$La-NQR spectra for LaPt$_{2-x}$Ge$_{2+x}$ ($x$ = 0 and 0.20).
For $x$ = 0, three clear peaks at $f$ = 2.16, 2.47, and 3.38 MHz are observed, which can be excellently understood with the parameters $\nu_{\rm Q} = 1.19 \pm 0.01$ MHz and $\eta = 0.73 \pm 0.01$.
The assignment of the each peak to the detailed transition is shown in the figure.
For $x$ = 0.20, two peaks at $f$ = 3.7 and 5.8 MHz, and a hump at $f$ = 2.9 MHz are observed.
The parameters $\nu_{\rm Q} = 1.97 \pm 0.08$ MHz and $\eta = 0.46 \pm 0.08$ can explain the data well, in which the two peaks are the transitions of $\pm 3/2 \leftrightarrow \pm 5/2$ and $\pm 5/2 \leftrightarrow \pm 7/2$, and the hump at 2.9 MHz is the $\pm 1/2 \leftrightarrow \pm 3/2$ transition.
Such assignment is consistent with the relative intensity of the three transitions as seen in $x$ = 0, and the $\nu_{\rm Q}$ value is consistent with that obtained from the $^{139}$La-NMR spectrum.
Figure \ref{NQR}(b) shows the recovery curves of the nuclear magnetization for $x$ = 0.20 measured at the $m = \pm 3/2 \leftrightarrow \pm 5/2$ transition.
As shown in Fig. \ref{NQR}(c), the $1/T_1$ is proportional to $T$ above $T_{\rm c}$, which is consistent with the $^{139}$La-NMR measurement.
Just below $T_{\rm c}$, the $1/T_1$ showed a small Hebel-Slichter peak and then decreased rapidly.
Thus we concluded that the superconducting gap of the $x$ = 0.20 sample is isotropically opened.

\section{Discussion}
The relationship between CDW and superconductivity has been investigated in systems by intercalation (Cu$_x$TiSe$_2$\cite{Morosan}, Cu$_x$TaS$_2$\cite{Wagner}), substitution (Lu$_5$Ir$_4$(Si$_{1-x}$Ge$_x$)$_{10}$\cite{Singh}) or applying pressure (2H-NbSe$_2$\cite{Berthier}, NbSe$_3$\cite{Briggs}).
The general feature of these systems is that suppressing CDW results in increasing $T_{\rm c}$.
The phase diagram of LaPt$_{2-x}$Ge$_{2+x}$ also has this feature.
It is natural to think that the $T_{\rm c}$ increases largely because of the increase in $N(E_{\rm F})$ by suppressing SPT.
To verify this idea, we used a rough approximation by fixing the Debye temperature $\theta_D$ and the attractive interaction $V$ to estimate the increase in $T_{\rm c}$  from $x$ = 0 ($T_{\rm c}$ =0.41 K).
We used McMillan formula\cite{McMillan} for this estimation,
\begin{equation}
T_{\text {c}}=\frac{\Theta _{\text {D}}}{1.45} \exp\left(-\frac{1.04 \left(1-\lambda\right)}{\lambda-\mu ^{*} \left(1-0.62\lambda \right)}\right)
\end{equation}
with $\theta_D$ = 310 K\cite{Tonohiro}, $\mu^\ast$ = 0.13\cite{McMillan} and $\lambda = VN(E_{\rm F})$, where $N(E_{\rm F})$ was obtained by band calculations.
As a result, we obtained $T_{\rm c}$ = 1.47 K for the tetragonal phase.
This value is smaller than $T_{\rm c}$ = 1.95 K obtained for $x$ = 0.20, but accounts for the majority of the increase in $T_{\rm c}$.
In cuprates and Fe-based superconductors, the magnetic fluctuations or structural/orbital fluctuations are believed to play an important role in producing superconductivity.
From comparison between $^{195}$Pt- and $^{139}$La-NMR, moderate fluctuations due to SPT were found in the present system.
In addition, we found that the stronger the fluctuations are, the higher the $T_{\rm {c}}$. 
Whether such fluctuations contribute to the rest of the increase of $T_{\rm c}$ merits investigation in the future.

\section{Summary}
We performed band calculations  for LaPt$_{2}$Ge$_{2}$ and found 2D-like Fermi surfaces with partial nesting.
A reduction in the density of states in the monoclinic phase  was found in the calculation and confirmed  by $^{195}$Pt-NMR measurements. We suggest a CDW as a possible cause for the SPT.
We synthesized non-stoichiometric LaPt$_{2-x}$Ge$_{2+x}$ samples and performed electrical resistivity, ac-susceptibility, and $^{195}$Pt and $^{139}$La NMR/NQR measurements.
We found that the SPT transition temperature ($T_{\rm{s}}$) decreases with increasing $x$, and as a result, the superconducting transition temperature ($T_{\rm c}$) increases from 0.41 K to 1.95 K, the majority of which can be accounted for by the  change in $N(E_{\rm F})$ associated with the reduction of  $T_{\rm{s}}$. 
Finally, we found moderate fluctuations associated with the SPT whose role in promoting superconductivity is a topic of future study.

\begin{acknowledgments}
We thank S. Kawasaki for technical help.
This work was partially supported by the \textquotedblleft Topological Quantum Phenomena\textquotedblleft \ Grant-in Aid for Scientific Research on innovative Areas from MEXT of Japan (Grant No. 22103004).
\end{acknowledgments}


\end{document}